\documentclass[5p,twocolumn,authoryear,9pt]{elsarticle}

\usepackage[english]{babel}

\usepackage{amsmath}
\usepackage{graphicx}
\usepackage{array}
\usepackage{bbold}
\usepackage{siunitx}
\usepackage{multicol}
\usepackage{float}
\usepackage{lipsum}
\usepackage{nicematrix}
\usepackage{comment}
\usepackage{relsize}
\usepackage[colorlinks=true, allcolors=blue]{hyperref}

\newcommand{\mt}[1]{\mathrm{#1}}

\journal{arXiv}
\begin{document}
\begin{frontmatter}

\title{Feedback stabilization of a nanoparticle at the intensity minimum\\ of an optical double-well potential}

\author[TU]{Vojt\v{e}ch Mlyn\'{a}\v{r}}
\author[UniVie]{Salamb\^{o} Dago}
\author[UniVie]{Jakob Rieser}
\author[UniVie]{Mario A. Ciampini}
\author[UniVie,IQOQI]{Markus Aspelmeyer}
\author[UniVie]{Nikolai Kiesel}
\author[TU,AIT]{Andreas Kugi}
\author[TU]{Andreas Deutschmann-Olek}

\affiliation[TU]{organization={Automation and Control Institute (ACIN), TU Wien},
            addressline={Gußhausstraße~27-29}, 
            city={Vienna},
            postcode={1040},
            country={Austria}}
            
\affiliation[UniVie]{organization={University of Vienna, Faculty of Physics, Vienna Center for Quantum Science and Technology (VCQ)},
            addressline={Boltzmanngasse~5}, 
            city={Vienna},
            postcode={1090},
            country={Austria}}

\affiliation[IQOQI]{organization={Institute for Quantum Optics and Quantum Information (IQOQI) Vienna, Austrian Academy of Sciences},
            addressline={Boltzmanngasse~3}, 
            city={Vienna},
            postcode={1090},
            country={Austria}}

\affiliation[AIT]{organization={AIT Austrian Institute of Technology},
            addressline={Giefinggasse~4}, 
            city={Vienna},
            postcode={1210},
            country={Austria}}

\begin{abstract}
In this work, we develop and analyze adaptive feedback control strategies to stabilize and confine a nanoparticle at the unstable intensity minimum of an optical double-well potential. The resulting stochastic optimal control problem for a noise-driven mechanical particle in a nonlinear optical potential must account for unavoidable experimental imperfections such as measurement nonlinearities and slow drifts of the optical setup. To address these issues, we simplify the model in the vicinity of the unstable equilibrium and employ indirect adaptive control techniques to dynamically follow changes in the potential landscape. Our approach leads to a simple and efficient Linear Quadratic Gaussian (LQG) controller that can be implemented on fast and cost-effective FPGAs, ensuring accessibility and reproducibility. We demonstrate that this strategy successfully tracks the intensity minimum and significantly reduces the nanoparticle’s residual state variance, effectively lowering its center-of-mass temperature.
While conventional optical traps rely on confining optical forces in the light field at the intensity maxima, trapping at intensity minima mitigates absorption heating, which is crucial for advanced quantum experiments.
Since LQG control naturally extends into the quantum regime, our results provide a promising pathway for future experiments on quantum state preparation beyond the current absorption heating limitation, like matter-wave interference and tests of the quantum-gravity interface.\\
\end{abstract}


\end{frontmatter}


\section{Introduction}\label{sec:introduction}
The levitation of mesoscopic particles is one of the most promising directions to push increasingly massive objects into the quantum regime \citep{gonzalez-ballestero_levitodynamics_2021}.
Traditional experiments use so-called traps to create a confining physical potential to keep the particle suspended in an (ultra-)high vacuum. Thereby, the trapping potential can be created by magnetic \citep{vinante_levitated_ferromagnetic_2020,timberlake_acceleration_sensing_2019}, electric \citep{paulElectromagneticTrapsCharged1990,bonvinHybridPaulopticalTrap2024} or optical \citep{ashkin_acceleration_1970} forces. The latter is exciting since it allows for optically measuring the particle's displacement with an accuracy close to the Heisenberg limit, which can be used to cool the motional degrees of freedom into the quantum ground state \citep{magrini_real-time_2021,tebbenjohanns_quantum_2021}. In optical traps, a nanoparticle is confined at the focal point of a strongly focused laser beam, where the intensity is maximal. Since the particle's refractive index is usually higher than that of the surrounding medium, it is naturally pulled towards the intensity maximum of the light field. Typically, this is achieved with silica spheres in air, up to ultra-high vacuum, and laser beams with a Gaussian transversal intensity profile (e.g., the fundamental TEM00 laser mode). However, exposing the particle to intense laser light results in internal heating of the trapped particle with several adverse effects: First, blackbody radiation from the heated surface induces decoherence, limiting the generation of non-classical quantum states. Second, it strongly limits the choice of materials due to evaporation in ultra-high vacuum, preventing the study of very promising hybrid materials. Finally, it hinders the effective use of cryostats, which are essential tools for working in the quantum regime. 

Keeping the particle in an intensity minimum of the light field, i.e., a dark spot, would thus be highly desirable. However, unless one submerges the particle into a medium with a higher refractive index \citep{almeidaTrappingMicroparticlesStructured2023}, dark spots are inherently unstable equilibrium points of the particle's dynamics, i.e., local maxima of the potential landscape. Remarkably, one can still infer the particle's position from light field measurements, although the particle is near a dark spot. 
Fundamentally, this enables a novel dark optical levitation scheme that relies on measurement information and feedback control algorithms to stabilize the particle at the unstable equilibrium point given by a dark spot.
In our recent work \citep{dago_stabilizing_2024}, we demonstrated a proof-of-principle experiment stabilizing a particle at the intensity minimum (apex) of a one-dimensional double-well potential. This work aims to present a comprehensive control-theoretic analysis and derive suitable (adaptive) control algorithms in detail. 

Hence, this paper gives an overview of the experimental setup in Section~\ref{sec:system_overview} and summarizes the mathematical model of the stochastic evolution of the particle and the measurements obtained from optical homodyne detection. In Section~\ref{sec:stab_strategy}, we derive different control algorithms to stabilize the particle at the minimum intensity and evaluate them in simulation studies. Finally, we present experimental results in Section~\ref{sec:experimental_results} and summarize our findings in Section~\ref{sec:conclusion}.

\section{System overview}\label{sec:system_overview}

\begin{figure}[!t]
    \centering
    \includegraphics[width=0.95\linewidth]{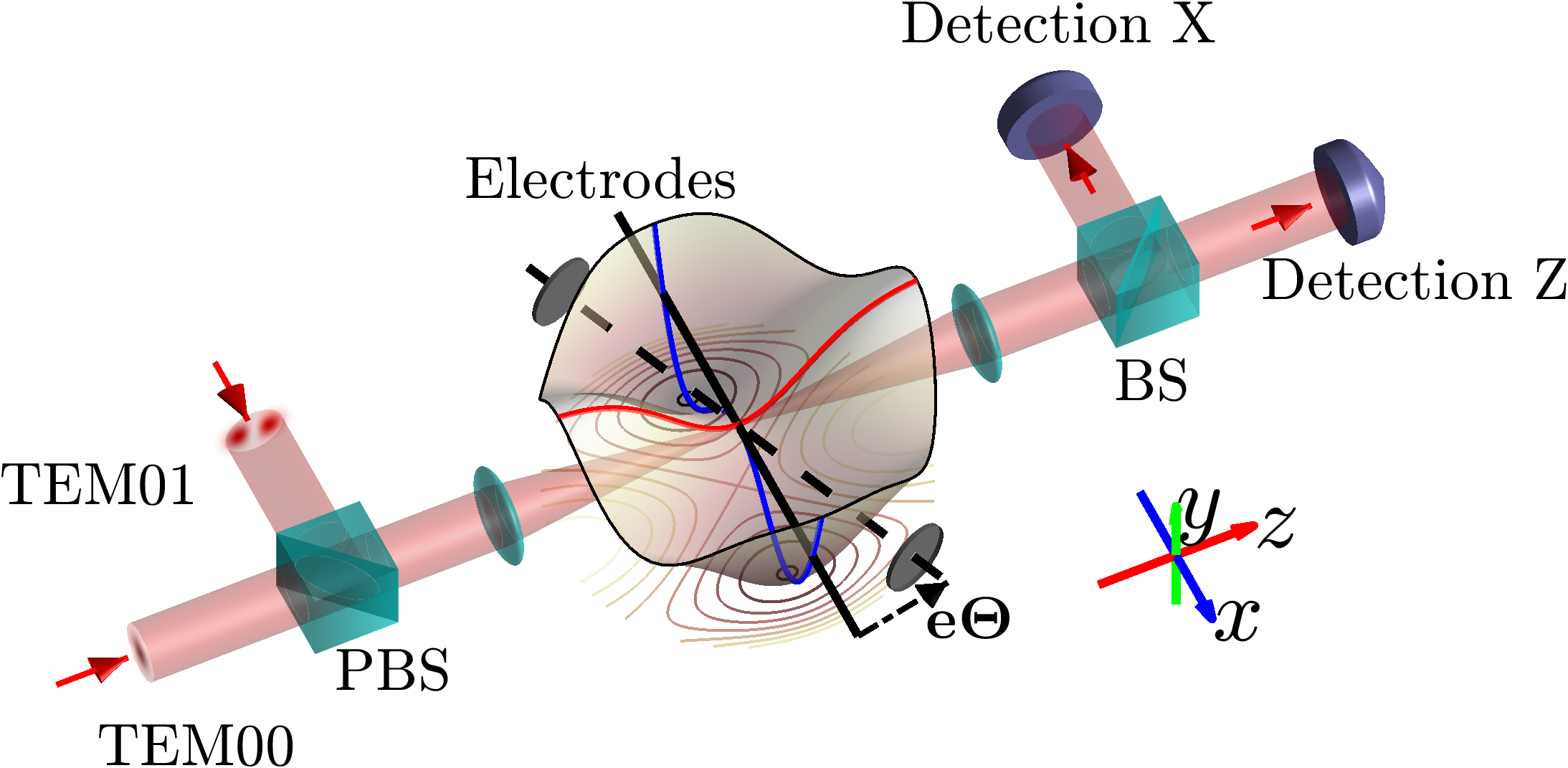}
    \caption{Simplified layout of the experimental setup. A TEM00 and a TEM01 beam are superimposed to create a one-dimensional double-well potential. Optical detection in the TEM00 beam lets one measure the particle's position in the $x$ and $z$ directions. The unstable $x$ direction is equipped with electrodes to act on the charged particle through electrostatic forces. The misalignment of the electrodes is exaggerated for illustration purposes.}
    \label{fig:simple_setup}
\end{figure}

In this work, we consider the experimental setup as shown in Fig.~\ref{fig:simple_setup}. A silica particle with a diameter of $210\pm5$~nm is optically trapped inside a vacuum chamber at low pressure (1~mbar) and room temperature. Due to collisions with residual gas molecules, the particle stochastically explores an optical potential constructed by superimposing two collinear, focused laser beams configured with transversal light-intensity modes TEM00 and TEM01. This configuration creates a tunable double-well potential in a direction perpendicular to the beam propagation and a (shallow) single-well potential in the remaining two degrees of freedom \citep{meloOpticalTrappingDark2020}. We define a coordinate system with $z$ corresponding to the direction of laser beam propagation and $x$ and $y$ being orthogonal to it. Furthermore, the $x$-axis is aligned along the double-well potential and is the primary degree of freedom in this paper.

After both laser beams pass through the vacuum chamber, the TEM00 beam is isolated and fed into two separate detectors: one for detecting the particle's position along the $x$-axis and the other for supplementary detection of the particle's position along the $z$-axis. However, it is important to note that both measurement channels are, to a certain extent, sensitive to particle displacements in the other axes.
To provide the ability to act on the particle along the double-well potential, electrodes are added inside the vacuum chamber on the $x$-axis. Applying a voltage to the electrodes allows us to exert an electrostatic force on the charged particles. However, imperfect alignment of the electrodes will result in small force components acting also on the $y$- and $z$-axes.
We refer the reader to \citep{dago_stabilizing_2024} for a more in-depth explanation of the experimental details. 

\subsection{Equations of motion}\label{subsec:equations_of_motion}
We assume that the particle is approximately spherical with homogeneous density and charge distribution on the surface. Hence, we consider only translational degrees of freedom $\mathbf{q}^\mt{T} = \left[x,\ y,\ z\right]$, leading to the Langevin equation \citep{gieselerNonequilibriumSteadyState2015}
\begin{equation} \label{eq:full_nlin}
m\ddot{\mathbf{q}}(t) + \gamma\dot{\mathbf{q}}(t) + \frac{\partial U(\mathbf{q},\mathbf{p})}{\partial \mathbf{q}} = \mathbf{c}_f u(t) + \bar{\mathbf{w}}(t) 
\end{equation}
with the mass of the particle $m$, the electrostatic force coefficients $\mathbf{c}_f^\mt{T} = \begin{bmatrix} c_{fx} & c_{fy} & c_{fz}\end{bmatrix}$, the voltage $u(t)$ applied to the electrodes, the damping coefficient $\gamma$, and the stochastic force $\bar{\mathbf{w}}^\mt{T}(t) = \begin{bmatrix} w_x(t) &  w_y(t) & w_z(t) \end{bmatrix}$. The latter two are a consequence of the interaction of the particle with the surrounding residual gas molecules at temperature \(T_0\). The stochastic force is thus given as Gaussian white noise
\begin{equation} \label{eq:force_noise}
\mathbb{E}[\bar{\mathbf{w}}(t),\bar{\mathbf{w}}^\mt{T}(t-\tau)] = 2\gamma k_B T_0 \mathbf{I} \,\delta(\tau) 
\end{equation}
with the identity matrix $\mathbf{I}$, the Boltzmann constant $k_B$, and the Dirac delta function $\delta(t)$.

As outlined above, the optical potential $U(\mathbf{q},\mathbf{p})$ with external parameters $\mathbf{p}$ originates from the interaction of a dielectric silica sphere with the two superimposed laser beams. The vector $\mathbf{p}^\mt{T} = \begin{bmatrix} P_{00} & P_{01} & \Delta_{0} & \Delta_{1} \end{bmatrix}$ contains those parameters of the actual potential shape that can be experimentally adjusted. These are, specifically, powers of both beams $P_{00/01}$ and the transversal offset of the beams in the $x$-axis $\Delta_{0/1}$ relative to the optical center-line due to imperfect alignment. These offsets are subject to slow drifts, which will be discussed in more detail later. The beams are formed as transversal Hermite-Gaussian light  modes, with their optical intensity distributions \citep{reiderPhotonics2016,novotnyPrinciplesNanoOptics2006} approximated as 
\begin{align} \label{eq:beam_intensity}
        I_{00} =& I_{00,0}(P_{00})\frac{ z_{00,0}^2}{z^2 + z_{00,0}^2}\times \\ &\qquad \exp{\left(\frac{-2y^2}{w_{00,y}^2(z)} - \frac{2\left(x-\Delta_0\right)^2 }{w_{00,x}^2(z)} \right)},\\
        I_{01} =& I_{01,0}(P_{01})\frac{ z_{01,0}^2}{z^2 + z_{01,0}^2} \frac{8\left(x-\Delta_1\right)^2}{w_{01,x}^2(z)}\times \\
        & \qquad \exp{\left(\frac{-2y^2}{w_{01,y}^2(z)} - \frac{2\left(x-\Delta_1\right)^2 }{w_{01,x}^2(z)} \right)}.
\end{align}

Here, $z_{00/01,0}$ are the Rayleigh lengths, and $w_{00/01,x/y}(z)$ are the local beam radii in the $x$- and $y$-axis along the light-propagation direction $z$. The waist is denoted in the two axes individually to take the asymmetry due to polarization into account. The strengths of the beams $I_{00,0}$ and $I_{01,0}$ are adjustable by tweaking the corresponding laser powers $P_{00}$ and $P_{01}$. Since the beams are orthogonally polarized and their phases are detuned, the combined intensity is easily obtained as a sum of their intensities. 
Hence, the optical potential for a dielectric particle under the Rayleigh approximation \citep{haradaRadiationForcesDielectric1996} is proportional to the intensity 
\begin{equation}\label{eq:pot_from_intensity}
U(\mathbf{q},\mathbf{p}) = \alpha'/4 \left[I_{00}(\mathbf{q},\mathbf{p}) + I_{01}(\mathbf{q},\mathbf{p})\right]
\end{equation}
with $\alpha'$ as the real component of the complex polarizability.

The powers of the beams drift over time, but slowly enough to be assumed constant during the experiment, and thus serve effectively as a configuration factor of the overall potential shape. The TEM00 beam pointing is spatially stabilized using piezo mirrors and a feedback loop before entering the vacuum chamber to mitigate low-frequency drift.  

\begin{figure}[!t]
    \centering
    \includegraphics[width=1.0\linewidth]{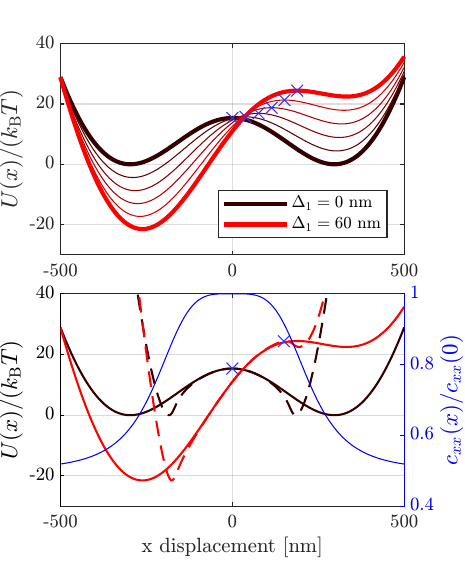}
    \caption{Potential landscape and nonlinear detection. \textbf{Top}: Changes of the potential due to the relative drift $\Delta_{1}$ of TEM01 to TEM00 (i.e., for $\Delta_0 = 0$ defining the reference frame of the optical detection). The black line depicts the potential for aligned beams, changing towards red for $\Delta_{1} =\SI{60}{\nano\meter}$. Blue crosses mark the apex (local maximum) of the optical potential. 
    \textbf{Bottom}: Normalized nonlinear detection sensitivity (blue) and resulting effective potential shapes (dashed) for aligned (black) and misaligned (red) cases.}
    \label{fig:potential_shapes}
\end{figure}

Furthermore, relatively slow drifts in both $\Delta_0$ and $\Delta_1$ occur on a time scale of approximately 100~ms. These are caused primarily by external environmental disturbances, and the beams are affected both simultaneously and independently due to differences in their propagation paths. The main observable effect is the potential shape change: When both beams drift together ($\Delta_0 = \Delta_1$), the potential moves relative to the detection reference frame, with the apex moving away from reference zero. When the beams also get misaligned relative to each other ($\Delta_0 \neq \Delta_1$), the apex moves away from zero, and the potential becomes imbalanced (see Fig.~\ref{fig:potential_shapes} Top).

\subsection{Optical detection scheme}\label{subsec:nlin_detection}
As mentioned previously, only the TEM00 beam is used for particle detection and is thus isolated from the orthogonally polarized TEM01 beam after passing the vacuum chamber. Then, a split self-homodyne detection and a photodetector with a pinhole give a voltage proportional to particle displacement in the $x$- and $z$-axes, respectively \citep{dago_stabilizing_2024}. 

The resulting measurement $\boldsymbol{\chi}(t) = \begin{bmatrix}\chi_x,& \chi_z \end{bmatrix}^\mt{T}$ can be described by
\begin{equation}\label{eq:full_nlin_detection}
\boldsymbol{\chi}(t) = \underbrace{\mathbf{h}(\mathbf{q})}_{\substack{\text{Nonlinear} \\ \text{output} \\ \text{function}}} + \underbrace{\mathbf{\Delta}_\chi(t)}_{\substack{\text{Reference} \\ \text{frame} \\ \text{drift}}} + \underbrace{\mathbf{v}(t)}_{\substack{\text{Measurement} \\ \text{noise}}}
\end{equation} with variable detection offset and measurement noise, respectively
\begin{equation}
\mathbf{\Delta}_\chi(t) = \begin{bmatrix} \Delta_{\chi,x}(t), \ 0 \end{bmatrix}^\mt{T} , \mathbf{v}(t) = \begin{bmatrix} v_{x}(t),\ v_{z}(t) \end{bmatrix}^T.
\end{equation}
For well-aligned detection and the particle in the near vicinity of the center of TEM00, the nonlinear output function $\mathbf{h}(\mathbf{q})$ can be approximated by a linear mapping
\begin{equation}\label{eq:detection_full_linear}
\mathbf{h}(\mathbf{q}) =
\begin{bmatrix} 
c_{xx} & c_{xy} & c_{xz}\\
c_{zx} &c_{zy} & c_{zz}
\end{bmatrix} 
\begin{bmatrix} 
x\\ y\\ z \end{bmatrix} 
\end{equation}
with $|c_{xx}| >> |c_{xy}|,|c_{xz}|$ and $|c_{zz}| >> |c_{zx}|,|c_{zy}|$. Nevertheless, both measurements exhibit significant crosstalk to other degrees of freedom. The coefficients $c_{ij}$, $i,j \in \{x,y,z\}$, are calibrated with the TEM01 beam turned off, that is, during a free evolution of the particle motion in a harmonic potential, and are assumed constant for the duration of the experiment and independent of $P_{01}$.

However, we observed a decrease in $c_{xx}$ coupled to the particle's displacement away from the TEM00 beam center on the $x$-axis. As illustrated in Fig.~\ref{fig:potential_shapes} (Bottom), there is a narrow, almost constant region of approx. $\pm200$~nm around the apex, with the detection sensitivity decreasing outside this region. This nonlinearity yields an apparent squashing of the potential observed in the measurement space, as illustrated in Fig.~\ref{fig:potential_shapes} (Bottom). This makes stabilizing the particle more challenging as the potential wells are closer to the apex in the measurement space.

Theoretically, we would not expect small drifts of the beam to affect the measurement of displacement along the $x$-axis, given the stabilization in place. However, mechanical drifts in optical components past the vacuum chamber add another component that effectively shifts the reference frame of the measurements along the $x$-axis. The resulting voltage offset $\Delta_{\chi,x}$ slowly changes at a rate of approximately
\begin{gather}
 \left| \frac{\mathrm{d}[\Delta_{\chi,x}(t)]}{\mathrm{d}t} \right| \leq \Delta_{\chi,x,max} \approx \SI{1e-4}{\volt\per\second},
\end{gather}
or, effectively, about $\SI{0.04}{\nano\meter\per\second}$, considering the typical detection sensitivity. This ultimately results in the loss of any reliable frame of reference. Since frequent realignment of the detection scheme is highly impractical, one has to resort to real-time tracking of the compound drift described in the following section.

\section{Control strategy}\label{sec:stab_strategy}

To prove the feasibility of levitation in the dark, as presented in Section \ref{sec:introduction}, we must ensure that the particle is kept at the minimum light intensity, that is, at the apex in the center of the double-well potential. Note that in this particular setup, the intensity minimum is not completely dark due to the TEM00 detection. However, the methods are directly transferable to a truly dark setup. 

Considering the drifts of the potential and detection described in Section \ref{sec:system_overview}, the goal of the controller is threefold: to stabilize the particle's center-of-mass dynamics, to suppress the stochastic effects of gas collisions, and to follow the slow-moving apex in an uncertain reference frame by only utilizing the noisy measurements $\mathbf{\chi}$ of \eqref{eq:full_nlin_detection}. Above all, however, the control strategy should avoid losing the particle.

Although our knowledge of the system is quite extensive in general, the compound drift of the main components, namely beam alignment $\Delta_0$ and $\Delta_1$ as well as detection drift $\Delta_{\chi,x}$, limits the usefulness of the nonlinear system description \eqref{eq:full_nlin} and \eqref{eq:full_nlin_detection}. In particular, separating and tracking its drifting components from available measurements to obtain estimates of $\mathbf{q}$ and $\mathbf{p}$ using a nonlinear observer is hampered by a lack of observability. 
Additionally, the drifts of the intensity profile at the particle plane cannot be measured directly inside or after passing through the vacuum chamber because they are indistinguishable from the effect of particle displacements. Hence, a different approach must be employed for their mitigation in the form of a tracking control scheme, allowing estimation of the actual position of the apex from the observed dynamics. Hence, the following section will simplify the model for tailored control design.

\subsection{Simplified model}\label{subsec:simplified_model}

\begin{figure}[t!]
    \centering
    \includegraphics[width=1.0\linewidth]{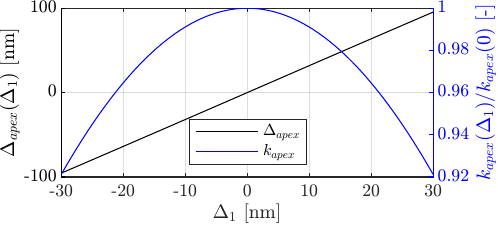}
    \caption{Behavior of the apex position $\Delta_{apex}(\mathbf{p})$ and the corresponding local stiffness $k_{apex}(\mathbf{p})$ for relative drift $\Delta_{1}$ of the two beams up to $\pm \SI{30}{\nano\meter}$.}
    \label{fig:xtip_ktip}
\end{figure}

First, we assume the particle remains sufficiently close to the center in the $y$ and $z$ directions.
Hence, the three degrees of freedom are approximately decoupled, and the potential is considered as a sum of individual components for each axis, specifically with a double well potential $U_x(x)$ along the $x$-axis and harmonic potentials along the $y$- and $z$-axes, i.e., 
\begin{equation}\label{eq:decoupled_pot}
    U \approx  U_x(x) + \frac{1}{2} m \Omega_y^2 y^2 + \frac{1}{2} m \Omega_z^2 z^2
\end{equation}
with the natural resonance frequencies $\Omega_y$ and $\Omega_z$, the mass of the particle $m$, and
\begin{equation}\label{eq:double_well_pot}
\begin{split}
    U_x = &\alpha(P_{00})\exp{\left(\frac{-2(x-\Delta_0)^2}{w_0^2}\right)}+ \\  \beta & (P_{01}) \left(\frac{x-\Delta_{1}}{w_0}\right)^2 \exp\left(\frac{-2(x-\Delta_{1})^2}{w_0^2}\right)
\end{split}  
\end{equation}
obtained from \eqref{eq:pot_from_intensity} by setting $y=0, z=0$, and assuming the beam waists $w_{00,x}(0) = w_{01,x}(0) = w_0$ are close to identical. The parameters $\alpha$ and $\beta$ are scaling factors determined by the geometry and physical properties of the experimental setup and are also tunable by the beam powers.

Since our goal is to stabilize the particle at the apex of the potential $\Delta_{apex}(\mathbf{p})$ given by
\begin{equation}
    x=\Delta_{apex}(\mathbf{p})\Leftrightarrow\frac{\partial U_x(x,\mathbf{p})}{\partial x}=0 \wedge \frac{\partial^2 U_x(x,\mathbf{p})}{\partial x^2} < 0,
\end{equation}
we can obtain a sufficiently accurate description in its vicinity by linearizing the potential along the $x$-axis, i.e.,
\begin{equation}\label{eq:Ux_simple}
U_{x} \approx \frac{1}{2} k_{apex}(\mathbf{p})\left(x - \Delta_{apex}(\mathbf{p}) \right)^2
\end{equation}
with 
\begin{equation} \label{eq:k_apex_definition}
    k_{apex}(\mathbf{p}) = \frac{\partial^2 U_x(x,\mathbf{p})}{\partial x^2} \bigg\rvert_{x=\Delta_{apex}(\mathbf{p})}.
\end{equation}

The changes of $\Delta_{apex}(\mathbf{p})$ and $k_{apex}(\mathbf{p})$ are exemplarily illustrated in Figure~\ref{fig:xtip_ktip} for relative drifts $\Delta_1$ up to 30nm, with $\Delta_0 = 0$, $P_{00} = \SI{80}{\milli \watt}$, and $P_{01} = \SI{135}{\milli \watt}$. 
The apex $\Delta_{apex}$ is displaced almost linearly relative to $\Delta_{1}$, while the local stiffness $k_{apex}$ is the highest for perfect alignment and decreases quadratically with $\Delta_{1}$, changing approximately 8\% at 30~nm.

Since the dynamics along the $y$-axis are stable and faster than the other two axes ($\tau_x \approx \SI{10}{\micro\second},\ \tau_y \approx \SI{6}{\micro\second}, \ \tau_z \approx \SI{28}{\micro\second}$, with $\tau_i = 1/\Omega_i$) and are only visible as crosstalk in the measurements, we entirely omit this degree of freedom. As we will show later, the $z$-axis cannot be neglected since its natural frequency is lower than that of the $x$-axis and is visible in both measurement channels. 

Hence, we replace the nonlinear potential with a simple approximation of an inverted quadratic potential of constant negative stiffness $k_{apex}$ and unknown offset of the center in the $x$-axis and a quadratic potential along the $z$-axis
\begin{equation}\label{eq:linpot}
U \approx \frac{1}{2} k_{apex}\left(x - \Delta_{apex} \right)^2 + \frac{1}{2}m\Omega_z^2 z^2.
\end{equation}

For the aligned and symmetric potential, the approximation in \eqref{eq:linpot} fits with an error within 1\% up to a 170~nm range, which is comparable to the extent of the linear detection region (see Section~\ref{subsec:nlin_detection}). This also means that as long as the beams remain sufficiently aligned and the controller succeeds in keeping the particle close to the apex, the quadratic approximation holds well and allows for controller designs that are substantially less demanding to implement on real-time hardware. 

Thus, using \eqref{eq:full_nlin} and \eqref{eq:linpot} while neglecting the motion in the $y$-direction we formulate the simplified model as 
\begin{subequations}\label{eqs:approx_statespace}
\begin{equation}
\dot{\boldsymbol{\xi}} = \mathbf{f}(\boldsymbol{\xi},\Delta_{apex}) + \mathbf{b}u +  \mathbf{G}\mathbf{w}\label{eq:approx_ss_1},
\end{equation}
with the vector $\boldsymbol{\xi}$ of state variables for displacement and velocity in $x$- and $z$-axes
\begin{equation}
\boldsymbol{\xi} = \begin{bmatrix} x & \dot{x} & z & \dot{z} \end{bmatrix}^T. 
\end{equation}
The function $\mathbf{f}$ is parametrized with the unknown apex drift $\Delta_{apex}$ according to
\begin{equation}
\mathbf{f}(\boldsymbol{\xi},\Delta_{apex}) = \begin{bmatrix}
  \dot{x} \\ -\frac{k_{apex}}{m}(x - \Delta_{apex}) -\Gamma \dot{x} \\
 \dot{z} \\ -\Omega_{z}^2 z -\Gamma\dot{z}
\end{bmatrix},
\end{equation}
where $\Gamma = \gamma/m$. The input vector $\mathbf{b}$ completes the deterministic part of the system as
\begin{equation}\label{eq:2d_input_vector}
\mathbf{b} = \begin{bmatrix} 0 & c_{fx}/{m} & 0 & c_{fz}/{m}\end{bmatrix}^T
\end{equation}
with force calibration factors $c_{fx}$ and $c_{fz}$, through which the electrode voltage $u$ acts on both axes. The force noises $\mathbf{w}$ stemming from the gas collisions enter the system via the process noise matrix $\mathbf{G}$
\begin{equation}
\mathbf{G} = \begin{bmatrix} 
0 & 0 \\
\frac{1}{m} & 0 \\
0 & 0 \\
0 & \frac{1}{m} \\
\end{bmatrix}, \quad
\mathbf{w} = \begin{bmatrix} w_x(t) \\ w_z(t)\end{bmatrix}
\end{equation}
\end{subequations}
and are assumed to be uncorrelated with zero-mean Gaussian probability distribution functions according to \eqref{eq:force_noise}.

To complete the state-space representation, we redefine the measured signals $\chi_{x}$ and $\chi_{z}$, neglecting crosstalk from the $y$-axis in \eqref{eq:full_nlin_detection}, \eqref{eq:detection_full_linear}, i.e.,
\begin{subequations} \label{eqs:approx_meas}
\begin{equation}
\boldsymbol{\chi} = \mathbf{C}\boldsymbol{\xi} + \mathbf{v} \label{eq:approx_meas_1}
\end{equation}
with
\begin{equation}
    \mathbf{C} = 
    \begin{bmatrix} 
        c_{xx} & 0 & c_{xz} & 0\\
        c_{zx} & 0 & c_{zz} & 0 
    \end{bmatrix}
\end{equation}
and Gaussian white measurement noise
\begin{equation}
\mathbb{E}\left[ \mathbf{v}(t)\mathbf{v}^\mt{T}(t-\tau) \right] =  \mathbf{R} \,\delta(\tau) = \mathrm{diag}\left( \begin{bmatrix} \sigma_x^2 & \sigma_z^2 \end{bmatrix} \right) \,\delta(\tau). 
\end{equation}
\end{subequations}
Figure \ref{fig:exp_psd} shows that the $x$-detection also contains the lower frequency of the $z$-axis motion component. In~that regard, $z$-detection serves primarily as an auxiliary source of information to improve the state estimation of the $z$ displacement and, conversely, to improve tracking of the potential's apex.

\subsection{Stochastically optimal control}\label{subsec:proposed_controller}
As mentioned before, the controller shall stabilize the particle at the unstable apex of the potential. Since its position is unknown, $\Delta_{apex}$ has to be estimated from available measurements. Furthermore, the particle is subject to strong stochastic excitations caused by collisions with residual gas molecules. These excitations can be suppressed by utilizing measurement information and feedback methods, also called feedback cooling, as this reduces the effective temperature of the mechanical degrees of freedom \citep{poggioFeedbackCoolingCantilevers2007, liMillikelvinCoolingOptically2011}. From a control perspective, this can be formulated as a stochastic optimal control problem that minimizes the quadratic cost function
\begin{equation}
J = \lim_{T \rightarrow \infty}  \mathbb{E} \left( \frac{1}{T}\int_{0}^{T} \boldsymbol{\xi}^\mathrm{T}_\mathrm{e} \mathbf{Q}_\mathrm{LQR}  \boldsymbol{\xi}_\mathrm{e} + r_\mathrm{LQR} u^2 \, \mt{d}t \right)
\end{equation}
with the weighting parameter $r_\mathrm{LQR} > 0$, 
\begin{equation}
\begin{split}
\mathbf{Q}_\mathrm{LQR} =& \mathrm{diag}\left(\begin{bmatrix} 
        \frac{\Omega_x}{2}, & \frac{\Omega_x}{2}, & q_z\frac{\Omega_z}{2}, & q_z\frac{\Omega_z}{2} 
    \end{bmatrix} \right), \\
    &\mathrm{with}\ \Omega_x = \sqrt{\left|k_{apex}\right|/m},
\end{split}
\end{equation}
the weighting parameter $q_z$, and the state of the control error
\begin{equation}\label{eq:true_state_vector}
    \boldsymbol{\xi}_e = \begin{bmatrix} x -\Delta_{apex} & \dot{x}-\dot{\Delta}_{apex} & z & \dot{z} \end{bmatrix}^T.
\end{equation}
From experimental observations, it is known that the apex position is drifting slowly compared to the timescale of the system dynamics \eqref{eqs:approx_statespace}. Thus, we assume that $\Delta_{apex}$ is unknown but constant for the controller design, i.e., $\dot{\Delta}_{apex}=0$.
From \eqref{eqs:approx_statespace}, one directly obtains the linear time-invariant dynamics of $\boldsymbol{\xi}_e$ according to
\begin{subequations} \label{eq:control_error_dynamics}
\begin{equation}
    \dot{\boldsymbol{\xi}}_\mathrm{e} = \mathbf{A}_e \boldsymbol{\xi}_\mathrm{e} + \mathbf{b} u + \mathbf{G} \mathbf{w}
\end{equation}
with 
\begin{equation}
    \mathbf{A}_e = \begin{bNiceArray}{cc:cc}
        0 & 1  & \Block{2-2}<\Large>{\mathbf{0}} \\
        -\frac{k_{apex}}{m} & -\Gamma  & \\
        \hdottedline
        \Block{2-2}<\Large>{\mathbf{0}} & & 0 & 1 \\
        & & -\Omega_{z}^2 & -\Gamma
    \end{bNiceArray}.
\end{equation}
\end{subequations}
Since the linear system \eqref{eq:control_error_dynamics} is fully controllable, there exists a unique solution of the (infinite-horizon) stochastic linear quadratic regulator (LQR) that is given by the linear state feedback law
\begin{equation} \label{eq:LQR_law}
    u = \mathbf{k}^\mathrm{T} \boldsymbol{\xi}_e = \mathbf{k}^\mathrm{T} 
    \begin{bmatrix}
        x -\Delta_{apex} \\ \dot{x} \\ z \\ \dot{z} 
    \end{bmatrix},
\end{equation}
where the optimal feedback gain $\mathbf{k}^\mathrm{T}$ is determined using the solution of the corresponding algebraic Riccati equation, see, e.g., \citep{ogataModernControlEngineering2010}. 

To implement the control law above, one needs to estimate the state $\boldsymbol{\xi}_e$, which is equivalent to estimating $\boldsymbol{\xi}$ and $\Delta_{apex}$. Therefore, we introduce the augmented state $\boldsymbol{\xi}_a = \begin{bmatrix} x & \dot{x} & \Delta_{apex} & z & \dot{z}  \end{bmatrix}^\mathrm{T}$ and consider slow changes of $\Delta_{apex}$ by assuming a random walk model
\begin{equation}\label{eq:random_walk}
    \dot{\Delta}_{apex}= w_{apex},
\end{equation}
where $w_{apex}$ as an independent white noise process with $\mathbb{E}\left[ w_{apex}(t), w_{apex}(t-\tau) \right] = \sigma_{w,apex}^2 \,\delta(\tau)$. It should be noted that $\sigma_{w,apex} \approx \SI{1e-6}{\meter^2 \second^{-2}}$ serves as a tuning factor to capture the drift of the apex and not deteriorate the estimates of the mechanical states.

The dynamics of $\boldsymbol{\xi}_a$ and the corresponding measurements are thus given by
\begin{subequations}\label{eq:system_for_KF}
\begin{align}
    \dot{\boldsymbol{\xi}}_\mathrm{a} &= \mathbf{A}_a \boldsymbol{\xi}_\mathrm{a} + \mathbf{b}_a u + \mathbf{G}_a \mathbf{w}_a\\
    \boldsymbol{\chi} &= \mathbf{C}_a \boldsymbol{\xi}_\mathrm{a} + \mathbf{v}
\end{align}
with $\mathbf{w}_a = \begin{bmatrix} w_x & w_z & w_{apex} \end{bmatrix}^\mathrm{T}$ and
\begin{align}
    \mathbf{A}_a = \begin{bNiceArray}{ccc:cc}
        0 & 1 & 0 & \Block{3-2}<\Large>{\mathbf{0}} \\
        -\frac{k_{apex}}{m} & -\Gamma & \frac{k_{apex}}{m} & \\
        0 & 0 & 0 & \\
        \hdottedline
        \Block{2-3}<\Large>{\mathbf{0}} & & & 0 & 1 \\
        & & & -\Omega_{z}^2 & -\Gamma
    \end{bNiceArray}.
\end{align}
The voltage applied to the electrodes $u$ and the process noise $w_a$ enter the system through
\begin{align}
    \mathbf{b}_a = \begin{bmatrix}
        0 \\ \frac{c_{fx}}{m} \\ 0 \\ 0 \\ \frac{c_{fz}}{m}
    \end{bmatrix}, \quad \mathbf{G}_\mathrm{a} = \begin{bmatrix} 
        0 & 0 & 0 \\
        \frac{1}{m} & 0 & 0 \\
        0 & 0 & 1 \\
        0 & 0 & 0 \\
        0 & \frac{1}{m} & 0 
    \end{bmatrix},
\end{align}
and the output matrix reads as
\begin{equation}
    \mathbf{C}_a = 
    \begin{bmatrix} 
        c_{xx} & 0 & 0 & c_{xz} & 0\\
        c_{zx} & 0 & 0 & c_{zz} & 0 
    \end{bmatrix}.
\end{equation}
\end{subequations}

The resulting optimal state estimator for the system description \eqref{eq:system_for_KF} is given by the Kalman (-Bucy) filter
\begin{align}\label{eq:kalman_filter}
    \dot{\hat{\boldsymbol{\xi}}}_a& = \mathbf{A}_a \hat{\boldsymbol{\xi}}_a + \mathbf{b}_a u + \hat{\mathbf{L}} \left( \boldsymbol{\chi} - \mathbf{C}_a \hat{\boldsymbol{\xi}}_a\right),
\end{align}
where the Kalman gain $\hat{\mathbf{L}}$ is determined using the solution of the corresponding algebraic Riccati equation, see, e.g., \citep{brownIntroductionRandomSignals2012}. 

The accuracy of the simplified model presented in \eqref{eqs:approx_statespace} degrades significantly when the setup experiences large drifts. This degradation arises from strong optical detection nonlinearities and potential asymmetries, as visualized in Figure~\ref{fig:potential_shapes}. Consequently, the reliability of the estimated apex position is limited to a region around the reference zero, which has been determined experimentally to be approximately $\pm \SI{100}{\milli\volt}$ in the measurement space.
Since the primary objective is to avoid losing the particle, also under severe drifts, one has to prevent the apex \emph{estimate} $\hat{\Delta}_{apex}$ from diverging into these regimes, even at the expense of larger tracking errors of the apex position. Hence, we demand that the apex estimate fulfills the constraints 
\begin{equation} \label{eq:box_constraints}
-\hat{\Delta}_{apex,max} \leq \hat{\Delta}_{apex}(t) \leq \hat{\Delta}_{apex,max}.
\end{equation}

Constrained state estimation problems are inherently nonlinear, and a large number of proposed methods with vastly different complexities exist in the literature for their solution, e.g., see \citep{KalmanFilterGeneralizations2006, simon_kalman_2010}.
Since the constraint above is only enforced on the estimated state, estimate projection strategies are an obvious choice.
Thereby, the unconstrained state estimate $\hat{\boldsymbol{\xi}}_a$ obtained from \eqref{eq:kalman_filter} is projected onto the feasible subset $\mathcal{S} \subset \mathbb{R}^5$ at every time step, which yields the projected state estimate $\tilde{\boldsymbol{\xi}}_a = \Pi_\mathcal{S}\left(\hat{\boldsymbol{\xi}}_a\right)$. The projection operator $\Pi_\mathcal{S}(\cdot)$ is given by the optimization problem
\begin{equation}
\tilde{\boldsymbol{\xi}}_a = \underset{\tilde{\boldsymbol{\xi}} \in \mathcal{S}}{\operatorname{arg min}} \| \tilde{\boldsymbol{\xi}}-\hat{\boldsymbol{\xi}}_a \|^2.
\end{equation}

For the box constraints \eqref{eq:box_constraints}, this projection simplifies to a straightforward clipping operation on $\hat{\Delta}_{apex}$, leaving the remaining states unaffected. Thus, the unconstrained estimate $\hat{\Delta}_{apex}$ obtained from solving \eqref{eq:kalman_filter} is simply truncated according to
\begin{equation}
\tilde{\Delta}_{apex} = \begin{cases} \hat{\Delta}_{apex,max} & \text{if } \hat{\Delta}_{apex} \geq  \hat{\Delta}_{apex,max}  \\
-\hat{\Delta}_{apex,max} & \text{if } \hat{\Delta}_{apex} \leq -\hat{\Delta}_{apex,max} \\
\hat{\Delta}_{apex} & \text{otherwise.}
\end{cases}
\end{equation}

The projected state estimate $\tilde{\boldsymbol{\xi}}_a$ can then be used in the control law \eqref{eq:LQR_law} to obtain the desired Linear Quadratic Gaussian (LQG) control algorithm. For the unconstrained case, the resulting closed control loop of the dynamics \eqref{eq:control_error_dynamics}, the feedback law \eqref{eq:LQR_law}, and the Kalman filter \eqref{eq:kalman_filter} is exponentially stable using the separation theorem, which can be easily seen since $\boldsymbol{\xi}_e$ and $\Delta_{apex}$ are related to $\boldsymbol{\xi}_a$ by a simple linear state transform. For the constrained case, the boundedness of $\boldsymbol{\xi}_e$ follows directly from $\Pi_\mathcal{S}(\cdot)$ being a non-expanding, continuous mapping, i.e., $\| \Pi_\mathcal{S}(\mathbf{x}) - \Pi_\mathcal{S}(\mathbf{y}) \| \leq \| \mathbf{x} - \mathbf{y}\|$. Indeed, this is always the case for convex sets $\mathcal{S}$, see, e.g., \citep{bauschke_convex_2011}.

\subsection{Simulation results}
{\begin{figure*}[t]
        \centering
        \includegraphics[width=1.0\textwidth]{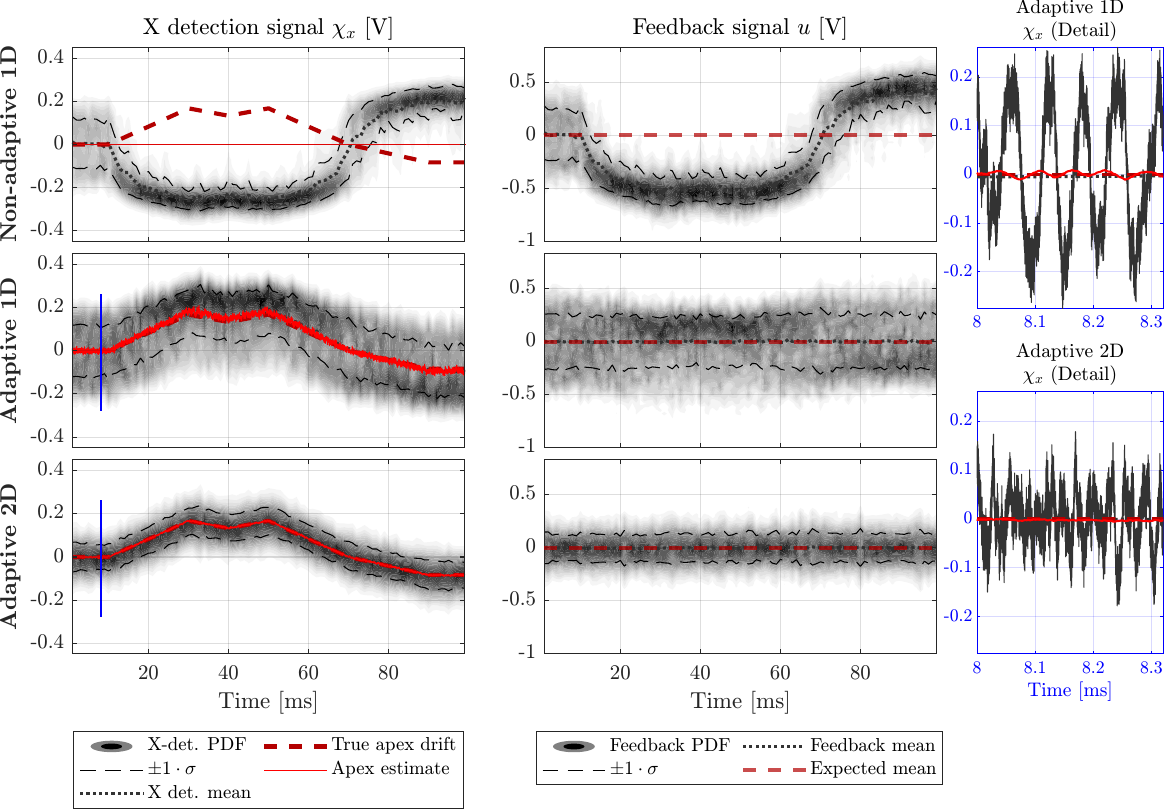}
        \caption{Qualitative comparison of the controller performance in simulation: All scenarios are initialized with identical noise realizations and initial conditions. As expected, including the estimate of $\Delta_{apex}$ into the Kalman Filter allows the controller to follow the drift of the local maximum. However, omitting the $z$-axis (\textit{Adaptive 1D} controller) leads to a significantly worse confinement of the particle (i.e., cooling). As illustrated in the detailed plot of the 8~ms mark on the right-hand side, this is due to the apex estimator wrongfully attributing cross-talk of $z$-axis motion in the $x$-detection signal.}
        \label{fig:sim_traces}
\end{figure*}}

\begin{figure}[t]
        \centering
        \includegraphics[width=1.0\linewidth]{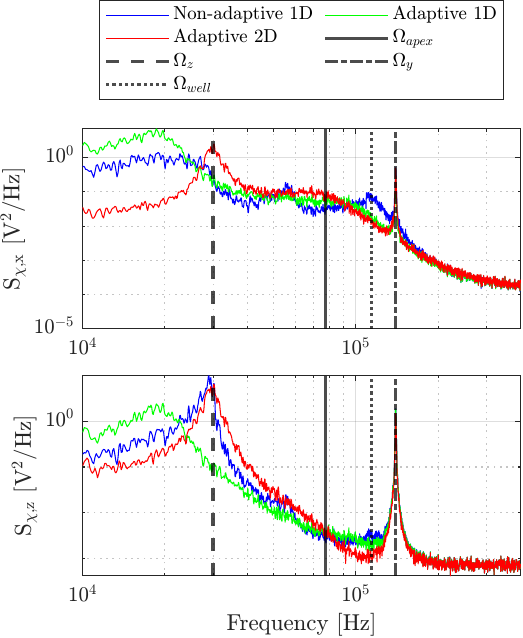}
        \caption{Power spectral densities of the two measurements for different simulated control algorithms: Non-adaptive 1D (blue), Adaptive 1D (green), and Adaptive 2D (red) for $x$-detection (top) and $z$-detection (bottom). The $\Omega_{well}$ peak in the non-adaptive controller indicates that the particle is exploring one of the potential wells.}
        \label{fig:sim_psd}
\end{figure}

To examine the controller performance, simulation studies were performed using the nonlinear plant model \eqref{eq:full_nlin} that includes the full optical potential \eqref{eq:pot_from_intensity} with a drift of TEM01 (i.e., $\Delta_1$) as well as detection nonlinearities according to \eqref{eq:full_nlin_detection}. The state feedback controller \eqref{eq:LQR_law} and the Kalman filter \eqref{eq:kalman_filter} are implemented with a fixed sampling frequency of $\SI{31.25}{\mega\hertz}$, including a total time delay of $ \SI{400}{\nano\second}$, mimicking real experimental conditions.

To motivate the inclusion of the $z$-axis and apex tracking in the estimator, we compare the indirect adaptive controller proposed in Section \ref{subsec:proposed_controller}, called \textit{Adaptive 2D} controller, with two simpler variants, \textit{Adaptive 1D} and \textit{Non-adaptive 1D}. The former still employs apex tracking (hence \textit{Adaptive}) but omits the $z$-axis, leaving the state vector $\boldsymbol{\xi}_{a1D}$ only with the displacement and velocity in the $x$-axis along the apex estimate
\begin{equation}
    \boldsymbol{\xi}_{a1D} = \begin{bmatrix} x & \dot{x} & \Delta_{apex} \end{bmatrix}^T. 
\end{equation}
For the \textit{Non-adaptive 1D}, the drift of the apex is additionally left out, leaving the state vector as
\begin{equation}
    \boldsymbol{\xi}_{1D} = \begin{bmatrix} x & \dot{x} \end{bmatrix}^T,
\end{equation}
which results in a simple LQG controller for a single harmonic oscillator.

Indeed, simple LQG controllers are the state-of-the-art approach for optimal feedback cooling for opto-mechanical systems, which usually considers only the primary axis while treating the transversal degrees of freedom as completely uncoupled. This assumption relies on the premise that the detection scheme is well-aligned and that the feedback action has a negligible effect on the other modes. Despite such simplifications, these algorithms are capable of feedback cooling the particle's center-of-mass motion into the quantum ground-state for Heisenberg-limited measurements \citep{magrini_real-time_2021}. 

To compare the performance of the controller variants, we are primarily interested in how they respond to the drift of the potential's apex $\Delta_{apex}$, which is observed to happen on a timescale of approximately $\SI{100}{\milli\second}$. The evolution of the probability distribution functions (PDF) of the measured detection signal $\chi_x$ and the corresponding feedback action $u$ are estimated through histograms of the recorded data over short time intervals of length $t_{avg}$.
Thereby, $t_{avg} = \SI{3}{\milli\second}$ is chosen significantly larger than the slowest time constant of the closed-loop system, which is typically the apex estimate (approx. $\SI{0.6}{\milli\second}$) of the Kalman filter. 

Figure \ref{fig:sim_traces} shows the time evolution of the PDFs for the proposed controller and its two alternative variants in simulation. Each variant faces the same scenario, starting with an ideally aligned potential ($\Delta_{0} = \Delta_{1} = 0$), with the TEM01 beam alignment $\Delta_{1}$ beginning to drift after the $\SI{10}{\milli\second}$ mark. As described in Section \ref{subsec:equations_of_motion}, this makes the potential asymmetric and shifts the apex of the potential ($\Delta_{apex} \propto \Delta_{1} \neq 0$).

As expected, the \textit{non-adaptive 1D} stabilizes the particle at the unstable equilibrium only as long as the apex is aligned with the reference zero. The corresponding  standard deviation of the tracking error is approximately $\SI{112}{\milli\volt}$ (or, equivalently, $\SI{42}{\nano\metre}$). As soon as the apex drifts away, the particle remains confined, but only on the slope of the potential. This is directly visible through the mean of the feedback action $u$ becoming non-zero, indicating that a static force is required to keep the particle at the steady-state position. Note that the apparent reduction in tracking error as the particle moves away from the apex is partially caused by the detection nonlinearity, see Fig.~\ref{fig:potential_shapes}. 

Conversely, both adaptive variants track the apex reliably, which is also confirmed by the zero-mean feedback action $u$, even when $\chi_x$ shows a non-zero mean behavior. However, the \textit{adaptive 1D} controller only achieves a significantly lower degree of confinement compared to the \textit{adaptive 2D} controller ($\SI{119}{\milli\volt}$ and $\SI{62}{\milli\volt}$ standard deviations, respectively) while requiring a higher control effort. Furthermore, the PDF of $\chi_x$ becomes asymmetric for high $\Delta_{01}$ between 20 and 60 ms. Both of these effects are caused by the Kalman filter of the \textit{adaptive 1D} controller wrongfully attributing the cross-talk of $z$-axis motion in the $x$-detection signal. This not only pollutes the apex estimate $\hat{\Delta}_{apex}$, see also the zoomed-in details on the right-hand side of Fig.~\ref{fig:sim_traces}, but also manipulates the dynamics in the $z$-direction as visible in the corresponding power spectral densities (PSDs) of the measurement signals in Fig.~\ref{fig:sim_psd}. Note that this is also true for the \textit{non-adaptive 1D} controller, where the spectral peak appears around $\Omega_{well}$, indicating that the particle is indeed exploring the potential well due to the misalignment of the potential. In summary, only the \textit{adaptive 2D} control algorithms, as developed in Section~\ref{sec:stab_strategy} can reliably confine the particle close to the apex of the potential.

\section{Experimental results} \label{sec:experimental_results}
Finally, the performance of the proposed controller is verified on an experimental setup, with a trapped \SI{210}{\nano\meter} diameter silica particle in a vacuum chamber at a pressure of \SI{1.5}{\milli\bar} and at room temperature. Section~\ref{sec:system_overview} already described the key ingredients of the setup from the control-engineering perspective. While the TEM00 beam is always active to obtain measurements, the TEM01 beam can be turned on and off during operation. In the absence of the TEM01 beam, the trapping potential effectively reduces to a simple harmonic potential, see Section~\ref{sec:system_overview}. For further details, we refer the reader to \citep{dago_stabilizing_2024}.

\subsection{Control performance criteria}\label{subsec:stab_crit}
To evaluate the performance of the control strategy in confining the particle at the potential apex, whose position is unknown, we formulate three criteria based only on statistics of the input-output data.  
\begin{description}
    \item[1. Unimodal PDF of $\chi_x$:] During free evolution, the particle frequently jumps between potential wells, leading to a distinct bimodal probability distribution function (PDF). For a successful controller, unimodality and a decrease in the variance of the PDF of the measured signal are expected.
    \item[2. Zero-mean feedback force:]  
    A particle at a local extremum of the potential implies that the feedback force is zero-mean, i.e., the feedback algorithm only has to counteract zero-mean stochastic forces. 
    In practice, this is considered satisfied when the average force is zero within $\pm1\sigma$ of its PDF.
    \item[3. Absence of spectral resonance peak:] 
    A particle confined at one of the potential wells instead of the apex would result in both, an unimodal PDF and a zero-mean feedback force. To discriminate this scenario, the particle motion must not exhibit oscillations at the resonance frequency of the potential wells, which is $\Omega_{well} = \SI{65}{\kilo\hertz}$ in our experimental setup.
    A prominent spectral peak near $\Omega_{well} = \SI{65}{\kilo\hertz}$ thus signifies the particle is indeed confined to one of the wells.
    
\end{description}
These criteria are also clearly visible in the simulation scenarios in Fig.~\ref{fig:sim_traces}: While all three controllers meet the first criterion (unlike a freely moving particle), the \textit{Non-adaptive 1D} controller does not meet the second criterion as soon as the apex drifts away from the reference zero. Then, the particle is kept at a slope instead, requiring a non-zero feedback force.

\subsection{Implementation and parameter calibration}
A configurable feedback controller is implemented at a sampling rate of \SI{31.25}{\mega\hertz} using a low-noise Red Pitaya STEMLAB 125-14 board containing a Zync 7020 SoC. Since the system parameters do not vary over time, the steady-state solution of the Kalman gain $\hat{\mathbf{L}}$ and the feedback gain $\mathbf{k}^\mt{T}$ are implemented by solving the Ricatti equation offline. Finally, the asymptotic LQG from \eqref{eq:LQR_law} and \eqref{eq:kalman_filter} is discretized in time through standard methods (see, e.g., \citep{franklinDigitalControlDynamic1998,vanloanComputingIntegralsInvolving1978}) and implemented on the FPGA using the Vitis Model Composer.

Note that for FPGA implementations with fixed-point arithmetic, it is highly beneficial to normalize the state and control variables prior to time discretization so that all state variables lie within the same order of magnitude.  The FPGA allows us to reconfigure the controller based on the calibrated parameters and switch instantaneously between internal models for the stable and unstable regimes, regardless of whether the TEM01 beam is active.

Most of the model parameters are obtained by fitting a power spectrum of a free particle evolution in a harmonic potential, i.e., with the TEM01 beam switched off. To calibrate $k_{apex}$, the TEM01 beam is switched on to create the double-well potential. By fitting the measured PDF of the particle's position in the $x$-direction to the expected PDF for a potential \eqref{eq:double_well_pot}, $k_{apex}$ is directly given using \eqref{eq:k_apex_definition}.
Finally, to calibrate the force calibration factors $c_{fx}$ and the $c_{fz}$ from \eqref{eq:2d_input_vector}, sine waves of known frequencies are applied to the electrodes, and then the height of the spectral peaks in the spectra of $\chi_x$ and $\chi_z$ is measured. The resulting parameters are listed in Table \ref{tab:calibrated_pars}. 

\begin{table}[h] \label{tab:calibrated_pars}
    \centering
    
    \begin{tabular}{l c}
        \hline
        \multicolumn{2}{c}{\textbf{Oscillator parameters}} \\
        \hline
        $\Gamma = \gamma/m$ &  $\SI{660}{\hertz}$ \\
        $\sqrt{|k_{apex}|/m}$ & $\SI{50}{\kilo\hertz}$ \\
        $\Omega_{z}$ & $\SI{46}{\kilo\hertz}$ \\
        $\Omega_{y}$ & $\SI{159}{\kilo\hertz}$ \\
        $\Omega_{x,well}$ & $\SI{65}{\kilo\hertz}$ \\
        \hline
        \multicolumn{2}{c}{\textbf{Force calibration factors}} \\
        \hline
        $c_{fx}$ & $\SI{-5.1e-13}{\newton\per\volt}$ \\
        $c_{fz}$ & $\SI{1.4e-13}{\newton\per\volt}$ \\
        \hline
        \multicolumn{2}{c}{\textbf{Detection calibration factors}} \\
        \hline
        $c_{xx}$ & $\SI{2.7e6}{\volt\per\meter}$ \\
        $c_{xz}$ & $\SI{0}{\volt\per\meter}$ \\
        $c_{zx}$ & $\SI{7.7e4}{\volt\per\meter}$ \\
        $c_{zz}$ & $\SI{1.1e6}{\volt\per\meter}$ \\
        \hline
    \end{tabular}
    \caption{Calibrated model parameters.}
\end{table}

\subsection{Experimental feedback stabilization}
{\begin{figure}[ht!]
        \centering
        \includegraphics[width=1.0\linewidth]{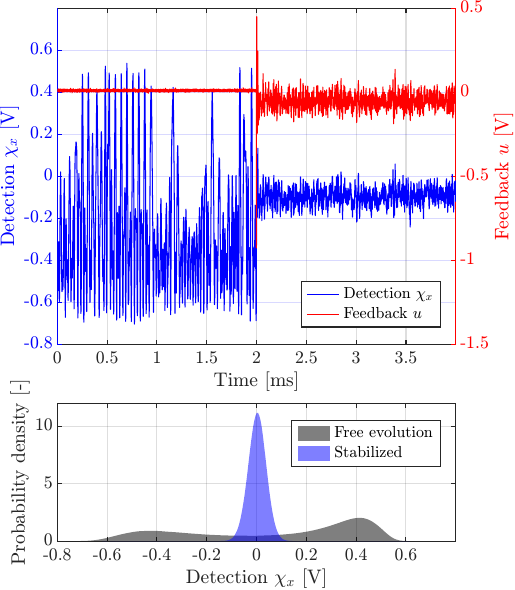}
        \caption{\textbf{Top:} Transient response when switching from the harmonic to the inverted potential configuration in the experiment (using the \textit{Non-adaptive 1D} controller). The controller is enabled at the 2~ms mark and quickly confines the particle close to the apex, indicated by the small mean force. \textbf{Bottom:} Probability density estimates of the $x$ detection signal $\chi_x$ before and after enabling the controller. The particle motion is visibly confined, as the PDF is turned unimodal. Note that the inverted potential was imbalanced during this run, which is why the side lobes of the PDF during free evolution are unequal in height.}
        \label{fig:exp_transient}
\end{figure}}
After a particle has been captured, the parameter calibration is completed, and the TEM01 beam is aligned with respect to the TEM00 beam, the particle is freely exploring the double-well potential. Activating the designed feedback algorithm at an arbitrary point, the particle is quickly confined after a brief transient period, as illustrated in Fig.~\ref{fig:exp_transient}.
Notably, this transition does not demand specific timing, even for the adaptive controller variants, underlying the control algorithm's reliability. This figure also illustrates the magnitude of the stochastic disturbances caused by gas collision, with the particle exploring almost the entirety of the potential landscape during free evolution. In contrast, enabling the feedback quickly suppresses the stochastic excitations and confines the particle to a significantly narrower position distribution. Also, notice the asymmetric PDF under free evolution due to an imbalanced double-well potential (see also Fig.~\ref{fig:potential_shapes}).

\subsection{Long-term behavior}
\begin{figure}[t]
        \centering
        \includegraphics[width=1.0\linewidth]{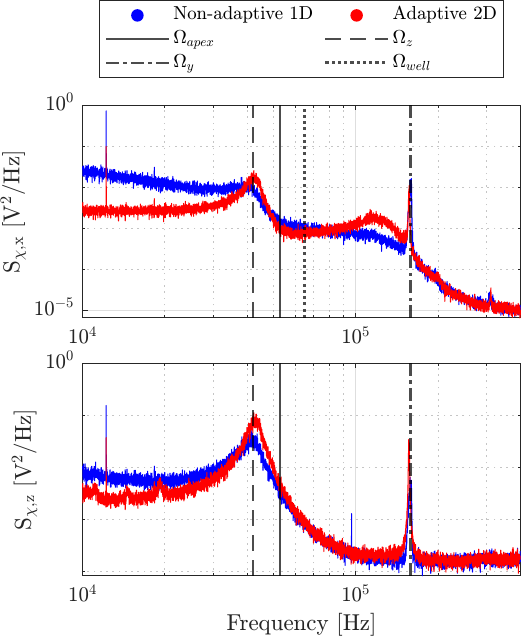}
        \caption{Power spectra from experimental data of $x$ detection $\chi_x$ (top) and $z$ detection $\chi_z$ (bottom) for the \textit{Non-adaptive 1D} (blue) and \textit{Adaptive 2D} (red) variants. The \textit{Adaptive 2D} variant attenuates the low-frequency components associated with the drifts, while both variants successfully keep the particle away from the potential wells, supported by the absence of a spectral peak at $\Omega_{well}$ in the $x$ detection $\chi_x$.}
        \label{fig:exp_psd}
\end{figure}

{\begin{figure*}[ht!]
        \centering
        \includegraphics[width=1.0\textwidth]{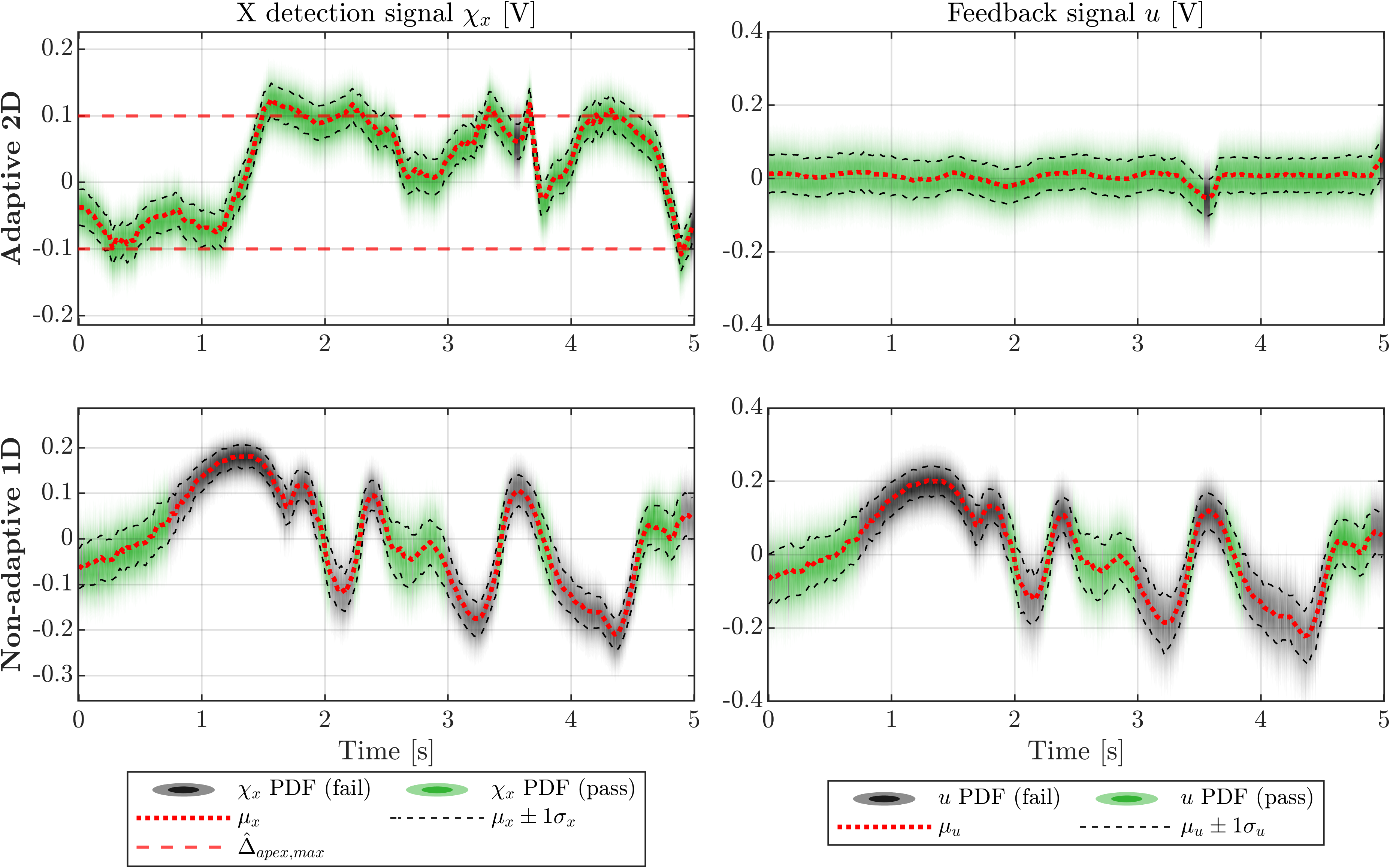}
        \caption{Comparison of two controller variants applied to the experimental setup: The time evolution of the probability density functions (PDFs) of $x$-detection (left) and feedback (right) signals, obtained by splitting the \SI{5}{\second} time trace into \SI{3}{\milli\second} intervals. Here, the \textit{Adaptive 2D} and \textit{Non-adaptive 1D} variants are compared, using time intervals with successful apex tracking according to the criteria in Sec.~\ref{subsec:stab_crit}, shaded in green.}
        \label{fig:exp_traces_compare}
\end{figure*}}

Finally, an experimental evaluation of the long-term performance of the proposed \textit{adaptive 2D} controller shall be presented. To this end, two representative experimental runs over \SI{5}{\second} each were chosen: One run where the experiment is sufficiently well-aligned such that the apex mostly remains within the chosen constraints (see Fig.~\ref{fig:exp_traces_compare}), and one where the misalignment is sufficiently strong such that the apex exceeds them (see Fig.~\ref{fig:exp_trace_tracking_fail}). In both plots, time intervals that meet all three performance criteria defined in Section \ref{subsec:stab_crit} are highlighted in green.

Figure~\ref{fig:exp_traces_compare} compares the proposed \textit{adaptive 2D} controller to the basic \textit{non-adaptive 1D} controller as a state-of-the-art reference. 
The \textit{adaptive 2D} controller successfully follows the drifting local maximum within the set constraints for the apex estimate (equivalent to $\pm$\SI{0.1}{\volt} in measurement space), thereby meeting the performance criteria outlined in the previous section. While the mean of the $x$-detection signal drifts wildly, the mean of the feedback is kept close to zero. As soon as the apex briefly reaches the constraint $\hat{\Delta}_{apex,max}$, such as around 3.6~s, the apex can no longer be tracked. The trajectory of the mean indicates that the probable cause is the apex moving past these limits, and thus, the particle remains on the slope instead. The tracking is restored when the apex drifts back within the set interval. 
In comparison, the \textit{non-adaptive 1D} controller meets the performance criteria only when the potential is accurately aligned, i.e., when the local maximum is at \SI{0}{\volt} in the reference frame. As the time trace shows, this configuration is hardly present and depends purely on chance because of external environmental factors, such as external temperature fluctuations, mechanical vibrations of various sources, etc. This is also reflected in the experimental spectra in Fig.~\ref{fig:exp_psd}, where the \textit{adaptive 2D} controller attenuates the low-frequency components compared to the \textit{non-adaptive 1D} controller. 

Figure~\ref{fig:exp_trace_tracking_fail} depicts the performance of the Adaptive 2D controller for the scenario with strong apex drifts that exceed the constraints of the apex estimate. For example, between the 2 and 3 second mark, the apex is apparently drifting below the lower limit of the apex estimate. Since the controller does not follow the apex further, the force of the potential pushes the particle into the opposite direction (cp. Fig.~\ref{fig:sim_traces}). As soon as the apex drifts back into the feasible region, tracking of the apex is restored until the next constraint violation happens. Ultimately, this experimental scenario highlights the robustness of the proposed control algorithm, avoiding the loss of a particle while tracking the apex as long as the experimental setup remains sufficiently aligned.

{\begin{figure*}[t]
        \centering
        \includegraphics[width=1.0\textwidth]{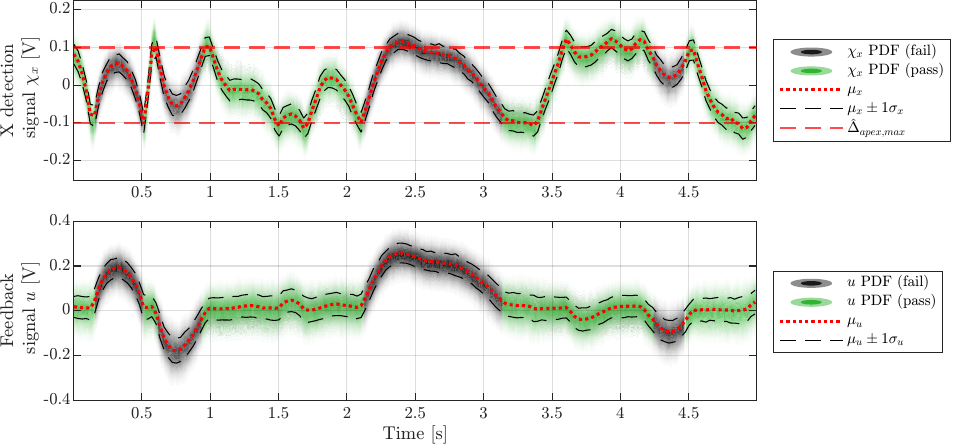}
        \caption{Experimental run showing a drift of the potential apex beyond the set limit $\hat{\Delta}_{apex,max}$ of the \textit{Adaptive 2D} controller. This is best demonstrated between 2 and 3 second marks, where the apex is assumed to have drifted below -0.1~V, with the controller attempting to counteract the additional resulting force caused by the particle being confined on the slope of the potential. After 2.4~s mark, the apex seemingly begins returning towards the region of confidence, decreasing the error until the tracking is restored until the next breach of the boundary. The same behavior can be observed between 0 and 1~s marks, with the apex exceeding both low and high limits in succession.
        \newline The time evolution of the probability density functions (PDFs) of $x$-detection (top) and feedback (bottom) signals, obtained by splitting the \SI{5}{\second} long time trace into \SI{3}{\milli\second} intervals. Time intervals with successful stabilization according to the criteria in Sec.~\ref{subsec:stab_crit} are shaded in green.}
        \label{fig:exp_trace_tracking_fail}
\end{figure*}}

\section{Conclusions and Outlook} \label{sec:conclusion}
In this paper, we developed and analyzed feedback control strategies to stabilize and cool the center-of-mass motion of a levitated nanoparticle at the intensity minimum of an optical double-well potential. Such control strategies are crucial for realizing optical dark traps that mitigate absorption heating while retaining the exceptional measurement accuracy of optical detection schemes.

This stabilization and cooling task essentially boils down to solving a stochastic optimal control problem for a mechanical system inside a nonlinear optical potential.
From a control engineering perspective, the primary challenge arises from the strong stochastic excitation coupled with relatively fast dynamics, experimental imperfections, including measurement nonlinearities, crosstalk in the electro-static actuation as well as the optical detection scheme, and slow drifts in the optical setup and laser power. These unavoidable drifts induce gradual changes in the optical potential and measurement signals, making it challenging to directly exploit knowledge of the nonlinear potential landscape. A linearization-based approach was adopted since these variations are not directly observable from the available measurements. This method captures the system dynamics in the vicinity of the apex of the potential (corresponding to the intensity minimum) and dynamically adapts to changes in the apex position over time. Further accounting for low-frequency transversal oscillations enhances tracking performance, significantly reducing the residual variance in the particle’s position error distribution and lowering its effective center-of-mass temperature.

The proposed stabilization method results in a simple Linear Quadratic Gaussian (LQG) control strategy, enabling real-time implementation on fast, cost-effective off-the-shelf FPGA boards like the Red Pitaya STEM-LAB SoC. Notably, the approach requires only parameters obtainable through standard calibration methods and does not rely on opaque tuning parameters, ensuring accessibility and reproducibility across similar experimental setups. Furthermore, since LQG optimal control naturally extends into the quantum regime \citep{edwardsOptimalQuantumFiltering2005}, it provides a promising framework for future experiments targeting the preparation of genuine quantum states of motion with a nanoparticle at a cold bulk temperature. This, in turn, would enable advanced matter-wave interference experiments \citep{batemanNearfieldInterferometryFreefalling2014, neumeierFastQuantumInterference2024} and contribute to the experimental investigation of the quantum-gravity interface \citep{aspelmeyerWhenZehMeets2022, rademacherQuantumSensingNanoparticles2020, fuchsMeasuringGravityMilligram2024}.

\section{Acknowledgments}
We thank Oriol-Romero-Isart and Patrick Mauerer for enlightening discussions on the FLIP project and Gregor Thalhammer for support on the optical setup. This research was funded in part by the Austrian Science Fund (FWF) [10.55776/COE1, PAT 9140723] and the European Union – NextGenerationEU. 
We acknowledge support from the Erwin Schrödinger Center for Quantum Science and Technology (ESQ) via a Discovery Grant.
This project has received funding from the European Research Council (ERC) under the European Union’s Horizon 2020 research and innovation programme (grant agreement No 951234, Q-Xtreme) and by the European Union (HORIZON TMA MSCA Postdoctoral Fellowships - European Fellowships, FLIP, No 101106514). Views and opinions expressed are, however, those of the author(s) only and do not necessarily reflect those of the European Union or the European Commission-EU. Neither the European Union nor the granting authority can be held responsible for them.

\bibliographystyle{elsarticle-harv} 
\bibliography{main}


\end{document}